\documentclass[11pt,a4paper]{article}
\usepackage{times}
\usepackage[T1]{fontenc}
\usepackage[utf8]{inputenc}
\usepackage[french]{babel}
\usepackage{fancyhdr}
\usepackage{setspace}
\usepackage{amssymb}
\usepackage{undertilde}
\usepackage{amsmath}
\usepackage{graphicx}

\usepackage{titlesec}
\titleformat{\section}
  {\normalfont\fontsize{16}{20}\bfseries}{\thesection}{1em}{}
\titleformat{\subsection}
  {\normalfont\fontsize{16}{20}\bfseries}{\thesubsection}{1em}{}
\titlespacing*{\section}
{0pt}{2.ex plus 1ex minus .2ex}{.0ex}
\titlespacing*{\subsection}
{0pt}{0.ex}{.0ex}

\begin{document}

\begin{center}
\begin{spacing}{2.05}
{\fontsize{20}{20}
\bf
Modélisation de la flexion d'un pseudo tricouche électro-actif à base de PEDOT, polymère semi-conducteur\\ }
\end{spacing}
\end{center}
\vspace{-1.25cm}

\begin{center}
{\fontsize{14}{20}
\bf
M. TIXIER\textsuperscript{a}, J. POUGET\textsuperscript{b}\\
\bigskip
}
{\fontsize{12}{20}
a. Université Paris-Saclay, UVSQ, CNRS, Laboratoire de Mathématiques de Versailles, UMR 8100, 78000, Versailles, France ; mireille.tixier@uvsq.fr\\
b. Sorbonne Université, CNRS, Institut Jean le Rond d'Alembert, UMR 7190, F-75005 Paris, France ; joel.pouget@upmc.fr\\}
\end{center}

\vspace{10pt}

{\fontsize{16}{20}
\bf
R\'esum\'e :}
\medskip

\textit{Les polymères électroactifs (PEA) sont des matériaux intelligents qui  peuvent être utilisés comme actionneurs, capteurs ou récupérateur d'énergie dans de nombreux domaines.
Nous nous sommes précédemment intéressés à des composites métal-polymère ioniques (IPMC), qui sont constitués d'un film de polymère ionique tel que le Nafion saturé d'eau et recouvert sur ses deux faces d'une fine couche de métal servant d'électrodes. Un tel système a la propriété de fléchir lorsqu'il est soumis à un champ électrique orthogonal au film et peut ainsi être utilisé comme actionneur. A l'inverse, le fléchissement du film génère une différence de potentiel entre les électrodes ; le même système peut donc être utilisé comme capteur.
Nous avons développé pour ce système un modèle de type "milieu continu". La thermodynamique des processus irréversibles linéaires nous avait permis d'établir ses lois de comportement.}

\textit{Nous nous intéressons actuellement à un système de propriétés voisines basé sur un PEA semi-conducteur, le PEDOT. La partie centrale du dispositif est constituée de deux polymères interpénétrés servant de réservoir d'ions. Le PEDOT est polymérisé de chaque côté et constitue un réseau interpénétré avec les deux autres polymères. L'ensemble forme un pseudo tricouche, les deux couches externes contenant le PEDOT faisant office d'électrodes. Il est ensuite saturé d'un liquide ionique. 
Lorsque la lame ainsi obtenue est placée dans un champ électrique orthogonal à ses faces, le PEDOT subit une réaction de réduction (ou dédopage) du côté de l'électrode négative, ce qui entraîne un afflux de cations provenant de la partie centrale et une augmentation de volume ; la lame fléchit donc vers l'électrode positive.}

\textit{Nous avons tout d'abord adapté notre modèle à ce système à deux constituants : les cations d'une part, et les trois polymères et les anions d'autre part. Nous avons écrit ses lois de conservation et ses relations thermodynamiques d'abord à l'échelle microscopique pour chaque constituant, puis à l'échelle macroscopique pour le matériau complet en utilisant une technique de moyenne. Nous en avons déduit, par la thermodynamique des processus irréversibles linéaires, ses lois de comportement : une loi rhéologique de type Kelvin - Voigt, une loi de Fourier et une loi de Darcy généralisées.
Les équations obtenues ont été appliquées au cas d'une lame encastrée - libre soumise à une différence de potentiel continue à température constante. La résolution numérique du système d'équations obtenu nous a permis de tracer la variation des différentes grandeurs dans l'épaisseur de la lame, qui sont des fonctions très raides au voisinage des extrémités. Nous avons également évalué la flèche et la force qu'il faut exercer sur l'extrémité libre de la poutre pour empêcher son déplacement (force de blocage). Les résultats obtenus sont en bon accord avec les mesures expérimentales publiées dans la littérature.}

\vspace{20pt}

{\fontsize{16}{20}
\bf
Abstract :}
\bigskip

\textit{Electroactive polymers (EAP) are smart materials that can be used as actuators, sensors or energy harvesters in many fields.
We had previously studied an ionic metal-polymer composites (IPMC), which consists in an ionic polymer film such as Nafion saturated with water and coated on both sides with a thin layer of metal acting as electrodes. This system bends when it is subject to an electric field orthogonal to the film and can thus be used as an actuator. Conversely, the deflection of the film generates a potential difference between the electrodes; the same system can therefore be used as a sensor.
We have developed a "continuous medium" model for this system. The thermodynamics of linear irreversible processes had enabled us to establish its constitutive equations.}

\textit{We are currently interested in a system of close properties based on PEDOT, a semiconductor EAP. The central part of the device consists in two interpenetrating polymers playing the role of an ions reservoir. The PEDOT is polymerized on each side and forms an interpenetrating network with the two other polymers. A pseudo trilayer is obtained, the two outer layers containing the PEDOT acting as electrodes. It is then saturated with an ionic liquid.
When the blade thus obtained is placed in an electric field orthogonal to its faces, the PEDOT undergoes a reduction reaction (or dedoping) on the side of the negative electrode, which attracts cations from the central part and therefore swells; the blade ultimately bends towards the positive electrode.}

\textit{We have first adapted our model to this two-components system: the cations on the one hand, and the three polymers and the anions on the other hand. We have written its balance equations and thermodynamic relations first at the microscopic scale for each phase, then at the macroscopic scale for the whole material using an averaging technique. The thermodynamics of linear irreversible processes then provides its constitutive relations: a Kelvin - Voigt type stress-strain relation and generalized Fourier's and Darcy's laws.
The equations obtained were applied to the case of a cantilevered blade subject to a continuous potential difference at constant temperature. The numerical resolution of the equations system enabled us to draw the profiles of the different quantities, which are very steep functions near the electrodes. We also evaluated the tip displacement and the force that must be exerted on the free end of the beam to prevent its displacement (blocking force). The results obtained are in good agreement with the experimental data published in the literature.}

\vspace{28pt}

{\fontsize{14}{20}
\bf
Mots clefs : Polymères électro-actifs - Couplages multiphysiques - Mécanique des polymères - PEDOT - Matériaux intelligents}

\section{Introduction}
\medskip
Les polymères électroactifs (PEA) sont des matériaux intelligents très prometteurs qui  peuvent être utilisés comme actionneurs ou capteurs dans des domaines variés : confection de micro-pompes, de micromanipulateurs ou de micro-robots, muscles artificiels, conception d'ailes battantes pour les micro-drones, récupération d'énergie....

Dans une étude précédente, nous nous sommes intéressés à des lames minces de polymères ioniques de type Nafion. Pour être actionnée, la lame doit être recouverte sur ses deux faces d'une fine couche de métal servant d'électrodes; l'ensemble forme un IPMC. La saturation de la lame en eau provoque une dissociation complète du polymère et la libération dans l'eau de cations de petite taille alors que les anions restent fixés sur les chaînes polymères. Ce système peut donc être modélisé comme un milieu poreux déformable dans lequel s'écoule un solvant (l'eau) et des cations, ces trois constituants ayant des champs de vitesses différents. Si l'on applique une différence de potentiel entre les deux électrodes, les cations migrent vers la cathode en entraînant avec eux l'eau par un phénomène d'osmose. Cette migration provoque un gonflement du polymère du côté de la cathode et une contraction sur l'autre face, entraînant une flexion de la lame vers l'anode. Ce processus met en jeu des couplages électro-mécano-chimiques que nous avons modélisés grâce à la thermomécanique des milieux continus. 
 
Nous avons ainsi obtenu les lois de comportement du système et les avons validées dans le cas statique en les comparant aux données expérimentales publiées dans la littérature \cite{Tixier1,Tixier2,Tixier3,Tixier4}.

Dans cet article, nous nous sommes intéressés au PEDOT (poly (3, 4 – éthylènedioxythiophène)), un PEA semi-conducteur. Pour l'utiliser dans la confection de capteurs ou d'actionneurs, il est nécessaire de l'associer à d'autres constituants jouant le rôle de réservoir d'ion. Le système que nous avons étudié a été optimisé par Festin et al \cite{Festin,Festin2013,festin2014}. Il est constitué d'une partie centrale constituée par deux polymères interpénétrés (IPN) : 60\% en masse de PEO (poly (oxyde d'éthylène)), un électrolyte solide, et 40\% de NBR (copolymère acrylonitrile - butadiène), un élastomère dont l'adjonction permet d'améliorer les caractéristiques mécaniques du mélange. Cette lame est ensuite plongée dans de l'EDOT, précurseur du PEDOT qui polymérise de chaque côté et constitue avec les deux autres polymères un réseau interpénétré. L'utilisation de réseaux interpénétrés de polymères permet d'éviter le délaminage. La proportion massique moyenne en PEDOT est de 18\%, mais il se concentre principalement sur les bords de la lame; la répartition du PEDOT dans l'épaisseur de la lame a été mesurée par Festin et al \cite{Festin2013}. L'ensemble forme un pseudo tricouche, les deux couches externes riches en PEDOT faisant office d'électrodes et la partie centrale de réservoir d'ions.

\begin{figure}[h]
\begin{center}
\includegraphics[height=2cm]{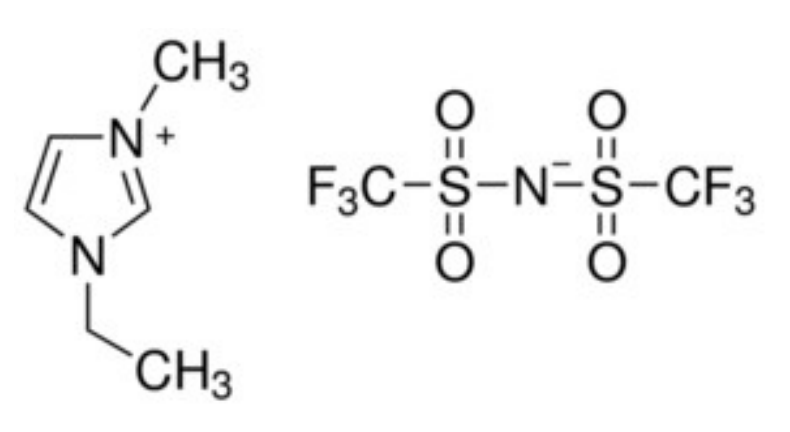}
\caption{EMITFSI}
\end{center}
\label{Form-EMI}
\end{figure}

L'ensemble est ensuite saturé d'un liquide ionique, l'EMITFSI (1-éthyl-3-méthyl-imidazolium bis (trifluorométhanesulfonyl) imide, figure \ref{Form-EMI}), qui pénètre presque exclusivement dans la partie centrale. La fraction massique moyenne de l'EMITFSI à saturation est de $57,3\%$ ; la composition finale de la lame est $13.6\%$ de NBR, $20.4\%$ de PEO et $7,7\%$ de PEDOT.

\begin{figure}[h]
\begin{center}
\includegraphics[height=3cm]{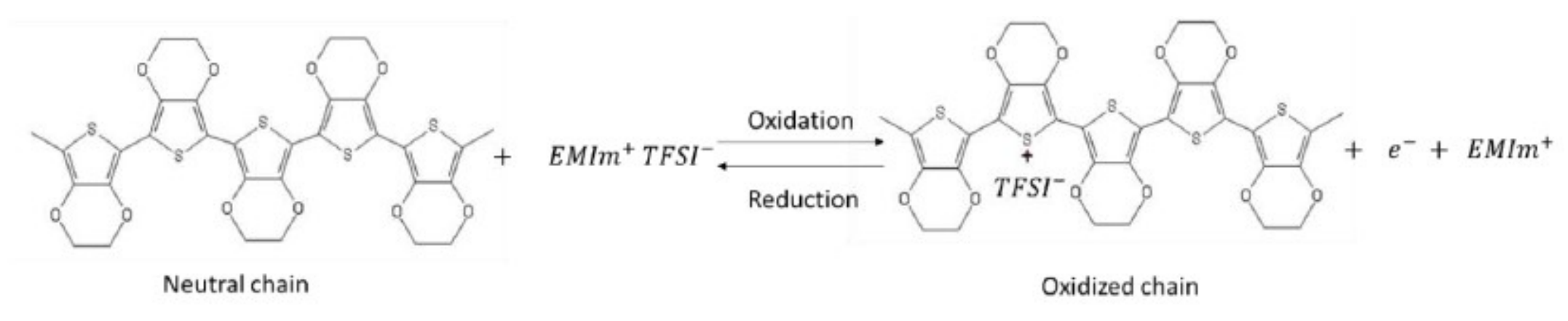}
\caption{Réactions d'oxydation / réduction d'un polymère conducteur \cite{nguyenTh}.}
\end{center}
\label{Dopage_PEDOT-EMI}
\end{figure}

L'EMITFSI est constitué d'anions $TFSI^{-}$ peu mobiles et de cations $EMI^{+}$. Lorsque le tricouche saturé est placé dans un champ électrique orthogonal à ses faces, le PEDOT subit une réaction de réduction (ou dédopage) du côté de l'électrode négative, ce qui entraîne un afflux de cations $EMI+$ provenant de la partie centrale et une augmentation de volume (figure \ref{Dopage_PEDOT-EMI}) ; la lame fléchit donc vers l'électrode positive. Les anions $TFSI-$, plus volumineux que les cations, restent insérés dans le réseau de polymères.

Ce système, comme les IPMC à base de Nafion, peut être modélisé grâce à la thermodynamique linéaire des processus irréversibles en utilisant une approche de type "milieu continu"; l'ensemble des trois polymères et des anions est assimilé à un milieu poreux dans lequel se déplacent les cations. 

L'adaptation de notre modèle au pseudo tricouche à base de PEDOT est présentée dans le paragraphe 2.
Dans le paragraphe 3, le modèle est appliqué à la flexion d'une lame encastrée à l'une de ses extrémités dans le cas statique et isotherme.
Le système obtenu est ensuite simplifié et adimensionné dans la quatrième partie.
La cinquième partie détaille les résultats de nos simulations numériques : les profils des différentes grandeurs dans l'épaisseur de la lame (potentiel, induction et charge électriques et pression). Nous avons également calculé la flèche et la force de blocage, c'est-à-dire la force que l'on doit exercer sur l'extrémité libre de la lame pour la maintenir immobile. Ces résultats sont discutés et comparés aux données expérimentales.
Nos conclusions sont exposées dans la dernière partie.

\section{Adaptation du modèle au pseudo tricouche à base de PEDOT}
\subsection{Modélisation du système et hypothèses}
\medskip
Nous avons modélisé le système par un milieu continu à deux phases : les trois polymères (PEO, NBR et PEDOT) et les anions $TFSI^{-}$ incrustés dans leurs chaînes sont assimilés à un milieu poreux solide déformable, homogène et isotrope se déplaçant avec le champ de vitesse $\overrightarrow{V_{2}}$; les cations $EMI^{+}$ constituent une phase liquide se déplaçant à la vitesse $\overrightarrow{V_{1}}$ dans les pores. Les deux phases sont séparées par une interface d'épaisseur nulle. Nous supposons en outre que la phase liquide est incompressible et que les déformations du solide sont petites. La gravité et l'induction magnétique sont négligeables.

Nous avons utilisé un modèle à gros grains développé pour les mélanges à deux constituants \cite{Ishii06}. Les équations de conservation sont écrites tout d'abord pour chacune des phases et pour les interfaces à l'échelle microscopique (typiquement $100 \; A^\circ$). Ces équations sont ensuite moyennées à l'échelle macroscopique (de l'ordre du $\mu m$) pour le matériau complet en utilisant une fonction de présence pour chacune des phases; il s'agit donc d'une moyenne spatiale, qui est supposée égale à une moyenne statistique (hypothèse d'ergodicité).

\subsection{Equations de bilan}
\medskip
On obtient ainsi les équations de bilan du matériau complet :

Conservation de la masse :%
\begin{equation}
\frac{\partial \rho }{\partial t}+div\left( \rho \overrightarrow{V}%
\right) =\frac{d\rho }{dt}+\rho div\overrightarrow{V}=0
\end{equation}
Equations de Maxwell, conservation de la charge électrique et relation constitutive :
\begin{equation}
\begin{tabular}{l}
$\overrightarrow{rot}\overrightarrow{E}=\overrightarrow{0}$ \\ 
$div\overrightarrow{D}=\rho Z$ \\ 
$div\overrightarrow{I}+\frac{\partial \left( \rho Z\right) }{\partial
t}=0$ \\ 
$\overrightarrow{D}=\varepsilon \overrightarrow{E}$%
\end{tabular}%
\end{equation}
Bilan de la quantité de mouvement :%
\begin{equation}
\rho \frac{D\overrightarrow{V}}{Dt}=div\utilde{\sigma }+\rho Z\overrightarrow{E}
\end{equation}%
où $\rho$ désigne la masse volumique, $\overrightarrow{V}$ la vitesse,  $\overrightarrow{E}$ le champ électrique, $\overrightarrow{D}$ l'induction électrique, $Z$ la charge électrique massique, $\overrightarrow{I}$ la densité volumique de courant, $\varepsilon$ la permittivité diélectrique et $\utilde{\sigma}$ le tenseur des contraintes. Les deux constituants ne se déplacent pas avec le même champ de vitesse ; afin de travailler avec un système fermé, nous avons introduit une dérivée "matérielle" $\frac{D}{Dt}$ ou dérivée en suivant le mouvement de chaque constituant, qui est une moyenne pondérée des dérivées particulaires  $\frac{d_{k}}{dt}$ relatives à chaque constituant \cite{Biot77,Coussy95} :
\begin{equation}
\rho \frac{D}{Dt}\left( \frac{g}{\rho }\right) 
=\sum\limits_{1,2,i} \rho _{k}\frac{d_{k}}{dt}\left( \frac{g_{k}}{\rho _{k}}\right) 
=\sum\limits_{1,2,i}\frac{\partial g_{k}}{\partial t}+div\left( g_{k}\overrightarrow{V_{k}}\right)
\end{equation}%
où $g$ est une grandeur volumique scalaire quelconque et $\rho_{k}$ la masse volumique de la phase $k$ rapportée au volume total du matériau; l'indice $1$ fait référence aux cations, $2$ à la phase solide, $i$ aux interfaces et les grandeurs sans indice sont relatives à l'ensemble des deux phases. Pour une grandeur vectorielle $\overrightarrow{g}$ :
\begin{equation}
\rho \frac{D}{Dt} \left( \frac{\overrightarrow{g}}{\rho }\right) %
=\sum\limits_{1,2,i}\frac{\partial \overrightarrow{g_{k}}}{\partial t}+div\left( \overrightarrow{g_{k}}\otimes \overrightarrow{V_{k}}\right) 
\end{equation}

On peut également écrire les bilans d'énergie cinétique $E_{c}$, potentielle $E_{p}$, interne $U$ et totale $E_{tot}=E_{c}+E_{p}+U$ et les échanges entre ces différents types d'énergies. Les équations sont résumées dans le tableau ci-dessous :
\begin{equation*}
\begin{tabular}{|l|l|l|l|l|l|}
\hline
Energie & &$Flux $&$E_{c}\longleftrightarrow E_{p}$ & $U\longleftrightarrow E_{p}$ & $E_{c}\longleftrightarrow U$ \\ \hline
$E_{p}$ &$\rho \frac{D}{Dt}\left( \frac{E_{p}}{\rho }\right) $&$=$& $- \rho Z \overrightarrow{V}.\overrightarrow{E}$ & $-%
\overrightarrow{i}.\overrightarrow{E}$ &  \\ \hline
$E_{c}$ &$\rho \frac{D}{Dt}\left( \frac{E_{c}}{\rho }\right)$&$=div\left(\utilde{\sigma}.\overrightarrow{V} \right)$& $+\rho Z \overrightarrow{V}.\overrightarrow{E}$ &  & $-\utilde{\sigma}:\utilde{grad}\overrightarrow{V}$ \\ \hline
$U$ &$\rho \frac{D}{Dt}\left( \frac{U}{\rho }\right) $&$=div\left[
\sum\limits_{1,2} \utilde{\sigma_{k}}.(\overrightarrow{V_{k}}-\overrightarrow{V})-\overrightarrow{Q'}\right]$ & & $+\overrightarrow{i}.\overrightarrow{E}$ & $+ \utilde{\sigma}:\utilde{grad}\overrightarrow{V}$ \\ \hline
$E_{tot}$ & $\rho \frac{D}{Dt}\left( \frac{E_{tot}}{\rho }\right)$&$ =div\left( \utilde{\sigma}.\overrightarrow{V} + \sum\limits_{k=1,2}\utilde{\sigma _{k}}.(\overrightarrow{V_{k}} - \overrightarrow{V}) -\overrightarrow{Q'} \right)$ & & &\\ \hline
\end{tabular}%
\end{equation*}
où $\overrightarrow{Q'}$ désigne le flux de chaleur par conduction et $\overrightarrow{i}=\overrightarrow{I} - \rho Z \overrightarrow{V}$ le courant électrique de diffusion.
Ces équations font apparaître les flux des différentes formes d'énergies (travail des forces de contact dans le référentiel barycentrique et flux de chaleur par conduction pour l'énergie interne, travail des forces de contact pour l'énergie cinétique) ainsi que les échanges entre les différents types d'énergie (termes sources). Ainsi le travail de la force électrique est un échange d'énergie potentielle et d'énergie cinétique, la chaleur dissipée par effet Joule une transformation d'énergie potentielle en énergie interne et la dissipation visqueuse une transformation d'énergie cinétique en énergie interne. On vérifie qu'il n'y a pas de terme source dans l'équation de bilan de l'énergie totale, qui est la somme des trois autres.

On peut enfin écrire l'équation de bilan de l'entropie volumique $S$ :
\begin{equation}
\rho \frac{D}{Dt}\left( \frac{S}{\rho }\right) =s-div\overrightarrow{\Sigma }
\end{equation}
où $s$ et $\overrightarrow{\Sigma}$ désignent respectivement la production volumique et le flux d'entropie.

\subsection{Relations thermodynamiques et lois de comportement}
\medskip
En faisant l'hypothèse de l'équilibre local, on peut écrire les relations de Gibbs, d'Euler et de Gibbs-Duhem du matériau \cite{deGroot} :
\begin{equation}
\begin{tabular}{ll}
$\rho T \frac{D}{Dt} \left( \frac{S}{\rho} \right)=\rho  \frac{D}{Dt} \left( \frac{U}{\rho} \right) +p \rho \frac{D}{Dt} \left( \frac{1}{\rho} \right)-\utilde{\sigma^{e}}^{s}:\utilde{grad}\overrightarrow{V}$
& Relation de Gibbs \\ 
$p=TS-U+\sum\limits_{k=1,2}\mu _{k}\rho _{k}$ & Relation d'Euler \\ 
$\phi _{1}\overrightarrow{grad}p=S_{1}\overrightarrow{grad}T+\rho _{1}\overrightarrow{grad}\mu_{1}$ & Gibbs-Duhem du fluide 1\\
$\phi _{2}\overrightarrow{grad}p=S_{2}\overrightarrow{grad}T+\rho _{2}\overrightarrow{grad}\mu _{2}- \sigma_{ij}^{es} \overrightarrow{grad}\epsilon_{ij}^{s}$ & Gibbs-Duhem du solide 2
\end{tabular}%
\end{equation}
où $T$ désigne la température absolue, $p$ la pression, $\utilde{\sigma^{e}}$ le tenseur des contraintes d'équilibre, $\mu_{k}$ et $\phi_{k}$ les potentiels chimiques massiques et les fractions volumiques des phases 1 et 2 et $\utilde{\epsilon}$ le tenseur des déformations; l'exposant $^{s}$ indique la partie symétrique de trace nulle d'un tenseur d'ordre 2.
En combinant la relation de Gibbs avec les équations de bilan de l'énergie interne et de la masse, on détermine la fonction de dissipation $s$ du système :
\begin{equation}
s=\frac{1}{T}\utilde{\sigma^{v}}:\utilde{grad}\overrightarrow{V}  +\frac{1}{T} \overrightarrow{E}.\overrightarrow{i} -\frac{1}{T^{2}} \overrightarrow{Q}.\overrightarrow{grad T}+\sum \limits_{1,2} \rho_{k} (\overrightarrow{V}-\overrightarrow{V_{k}}).\overrightarrow{grad} \left( \frac{\mu _{k}}{T} \right)
\end{equation}
où $\utilde{\sigma^{v}}=\utilde{\sigma}-\utilde{\sigma^{es}}-p\utilde{1}$ désigne le tenseur des contraintes dynamiques et $\overrightarrow{Q} = \overrightarrow{Q'}+ \sum\limits_{1,2} \left[ U_{k}(\overrightarrow{V_{k}} - \overrightarrow{V}) -\utilde{\sigma _{k}}.(\overrightarrow{V_{k}}-\overrightarrow{V}) \right]$ le flux de chaleur.
On peut alors identifier les flux et les forces généralisées associées:
\begin{equation}
\begin{tabular}{|l|l|}
\hline
Flux & Forces généralisées\\ \hline
$\frac{1}{3}tr\left( \utilde{\sigma ^{v}}\right)
\quad $ & $\frac{1}{T}div\overrightarrow{V}$ \\ \hline
$\overrightarrow{Q}$ & $\overrightarrow{grad}\left( \frac{1}{T}%
\right) $ \\ \hline
$\overrightarrow{J_{m}}= \rho_{1}(\overrightarrow{V_{1}}-\overrightarrow{V_{2}})$ & $\frac{\rho_{2}}{\rho} \left[ \frac{Z_{1}-Z_{2}}{T} \overrightarrow{E}+ \overrightarrow{grad} \left( \frac{\mu_{2}-\mu_{1}}{T} \right) \right]$ \\ \hline
$\utilde{\sigma ^{v}}^{s}$ & $\frac{1}{T}%
\utilde{grad}\overrightarrow{V}^{s}$ \\ \hline
\end{tabular}%
\end{equation}

La thermodynamique des processus irréversibles linéaires permet d'en déduire les lois de comportement du matériau. On obtient une loi de Fourier généralisée, une loi rhéologique de type Kelvin-Voigt:
\begin{equation}
\utilde{\sigma}=\frac{E \nu}{(1-2 \nu)(1 + \nu)} \left( tr\utilde{\epsilon}\right) \utilde{1}+\frac{E}{1 + \nu}\utilde{\epsilon}+\lambda _{v}\left( tr\overset{\bullet }{\utilde{\epsilon}}\right) \utilde{1}+2\mu _{v}\overset{\bullet }{\utilde{\epsilon}}
\end{equation}
et une loi de Darcy généralisée, qui s'écrit dans le cas isotherme :
\begin{equation}
\overrightarrow{V_{1}}-\overrightarrow{V_{2}}= -\frac{K}{\eta \phi_{1} } \left[ \overrightarrow{grad}p + \left( \frac{1} {\rho_{2}^{0}} - \frac{1}{\rho _{1}^{0}} \right)^{-1} \left((Z_{1}-Z_{2})\overrightarrow{E} +\frac{1}{\phi_{2} \rho_{2}^{0}}\sigma_{ij}^{es} \overrightarrow{grad}\epsilon_{ij}^{s} \right) \right]
\end{equation}
$E$ désigne le module d'Young, $\nu$ le coefficient de Poisson, $\lambda_{v}$ et $\mu_{v}$ les coefficients viscoélastiques, $\eta$ la viscosité dynamique de la phase liquide, $K$ la perméabilité absolue de la phase solide et $\rho_{k}^{0}$ la masse volumique de la phase $k$ ramenée à son volume; les $\overset{\bullet}{ }$ désignent des dérivées temporelles.

\section{Système d'équations pour une lame en flexion (cas statique)}
\subsection{Modélisation de la poutre en flexion}
\begin{figure}[h]
\begin{center}
\includegraphics[height=2.5cm]{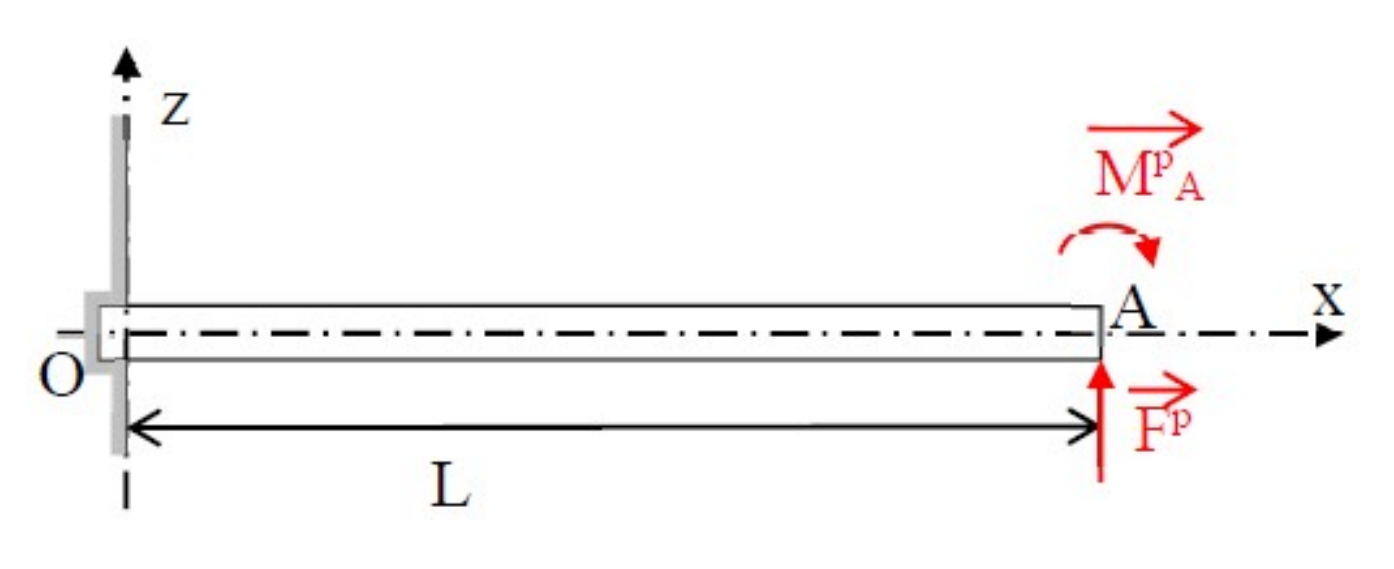}
\qquad 
\includegraphics[height=2.5cm]{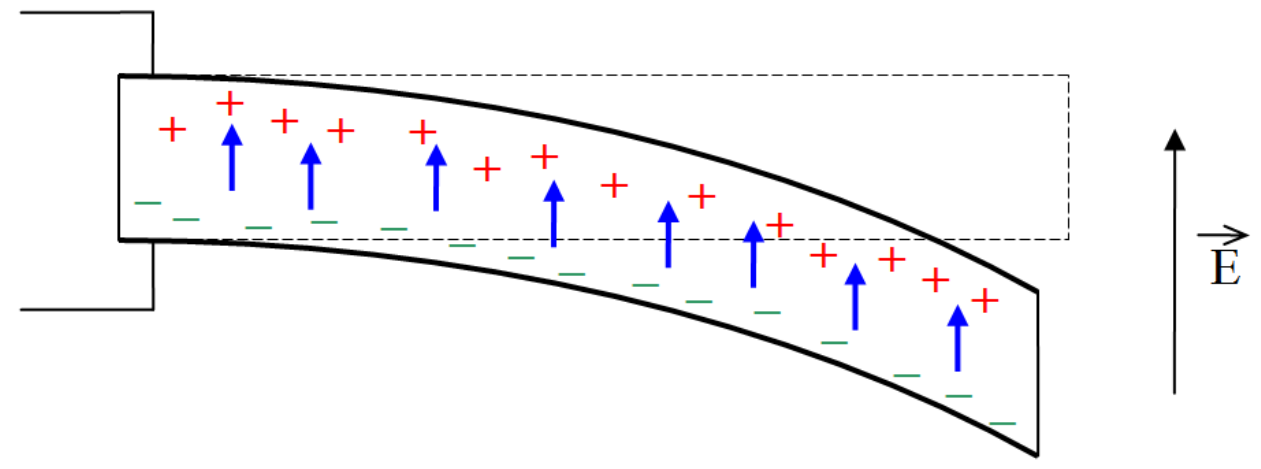}
\caption{Efforts exercés sur la poutre; lame de PEDOT en flexion}
\label{Poutre}
\end{center}
\end{figure}

\medskip
Nous avons appliqué ce modèle à la lame de PEDOT décrite dans l'introduction se déformant sous l'action d'un champ électrique permanent. Nous supposons que le phénomène est isotherme. Les dérivées partielles par rapport au temps, les gradients de température et les champs de vitesses des deux constituants sont donc nuls. La lame étudiée a une épaisseur $2e = 250 \mu m$, une largeur $2 \ell = 11 mm$ et une longueur $L=18 mm$.
La lame étant mince, on peut utiliser un modèle de poutre en petites déformations et petits déplacements pour la décrire. La poutre est encastrée à son extrémité $O$. L'autre extrémité $A$ est soit libre, soit soumise à un effort tranchant $\overrightarrow{F^{p}}$ bloquant son déplacement. Lorsqu'on applique une différence de potentiel $2 \varphi_{0} = 4 V$, les cations se déplacent vers l'électrode négative, entraînant une augmentation de volume qui provoque la flexion de la lame vers l'électrode positive (figure \ref{Poutre}).
Les forces appliquées à la poutre peuvent être modélisées par un moment de flexion $\overrightarrow{M^{p}}$ d'axe $Oy$ et par la force de blocage $\overrightarrow{F^{p}}$ appliqués en $A$.

On choisit un repère $Oxyz$ tel que l'axe $Ox$ soit suivant l'axe de la poutre non déformée, l'axe $Oz$ orthogonal à la lame et l'axe $Oy$ selon la largeur. Nous faisons les hypothèses habituelles : les sections droites restent planes et normales aux fibres après déformation (hypothèse de Bernoulli) et la répartition des contraintes est indépendante des points d'application des forces extérieures (hypothèse de Barré Saint Venant). Le moment fléchissant s'exprime en fonction du champ de pression $p=-\frac{1}{3}tr \utilde{\sigma^{e}} \simeq -\frac{\sigma_{xx}}{3}$ :
\begin{equation}
M^{p}= \int_{-l}^{l} \int_{-e}^{e}\sigma_{xx}~z~dz~dy = -6l \int_{-e}^{e} p~z~dz
\label{moment}
\end{equation}
La poutre est en flexion pure. On en déduit l'expression de la flèche $w$, de l'angle de rotation $\theta$ et de la déformation $\epsilon_{xx}$ dans le cas de la poutre encastrée libre ainsi que la force de blocage $\overrightarrow{F^{p}}$ :
\begin{equation}
\begin{tabular}{ll}
$w=-\frac{M^{p}}{2E_{moy}I^{p}}L^{2} \qquad \qquad $&$\theta = \frac{M^{p}}{E_{moy}I^{p}}L$ \\ 
$\epsilon_{xx} =\frac{M^{p}e}{E_{moy}I^{p}}$&$F^{p}=\frac{3}{2}\frac{M^{p}}{L}$ \\
\end{tabular}
\label{eq-meca}
\end{equation}
où $I^{p}$ désigne le moment quadratique de la poutre par rapport à l'axe $Oy$ :
\begin{equation}
I^{p}=\int_{-l}^{l}\int_{-e}^{e}z^{2}\;dz\;dy=\frac{4le^{3}}{3}
\end{equation}

\subsection{Equations pour la lame en flexion dans le cas statique}
\medskip
Le problème peut être considéré comme bidimensionnel dans le plan $Oxz$. On peut en outre considérer qu'en première approximation, les composantes $E_{x}$ et $D_{x}$ du champ et du déplacement électriques sont négligeables et que le champ, l'induction, la charge et le potentiel électriques $E_{z}$, $D_{z}$, $\varphi$ et $\rho Z$, la pression $p$ et la fraction volumique de cations $\phi_{1}$ ne dépendent que de la composante $z$. La variation relative de volume du matériau $tr \utilde{\epsilon}$ est par ailleurs due au mouvement des cations, donc reliée à $\phi_{1}$. Le tenseur des déformations est lié à l'expansion du volume $V$ du matériau :
\begin{equation}
tr \utilde{\epsilon} = \frac{\Delta V}{V} = \epsilon_{xx}
\end{equation}
Les variations de volume du matériau sont inférieures à $\epsilon_{xx,max}=2,8 \%$ \cite{festin2014} et sont donc négligeables. En supposant que le volume de la phase 2 est invariant au cours de la déformation, on en déduit :
\begin{equation}
tr \utilde{\epsilon} \simeq \phi_{1} - \phi_{1r}
\end{equation}
où $\phi_{1r}$ désigne la fraction volumique de la phase 1 dans la poutre non déformée. Le système d'équations s'écrit alors :
\begin{equation}
\begin{tabular}{l}
$E_{z}=-\frac{d \varphi}{dz}$ \\
$\frac{dD_{z}}{dz}= \phi _{1} (\rho _{1}^{0}Z_{1}-\rho_{2}^{0} Z_{2}) +\rho_{2}^{0} Z_{2} =\rho Z$ \\
$D_{z}=\varepsilon E_{z}$ \\
$\left( \frac{1} {\rho_{2}^{0}} - \frac{1}{\rho _{1}^{0}} \right) \frac{dp}{dz} + (Z_{1}-Z_{2})E_{z} +\frac{1}{\phi_{2}\rho_{2}^{0}}\frac{6(1+\nu)}{E} p \frac{dp}{dz}=0$ \\
$\phi_{1r} - \phi_{1}= \frac{3 (1 - 2 \nu)}{E} p$ \\
\end{tabular}
\end{equation}
avec les conditions aux limites :
\begin{equation}
\varphi(e) = - \varphi_{0} \qquad \qquad \varphi(-e) = \varphi_{0} \qquad \qquad D_{z}(e) = D_{z}(-e)
\end{equation}
Cette dernière condition, qui traduit l'électroneutralité, peut s'écrire, dans le cas où la permittivité est constante dans toute la lame, c'est-à-dire indépendante de la composition :
\begin{equation}
\left. \frac{d\varphi}{dz} \right)_{e} =\left. \frac{d\varphi}{dz} \right)_{-e}
\end{equation}

La masse volumique et la charge massique du liquide valent respectivement $\rho_{1}^{0}= 1,53~g~cm^{-3}$ et $Z_{1}=8,69~10^{5}~kg~m^{-3}$ et le coefficient de Poisson peut être évalué à $\nu \simeq 0,4$ (courtoisie de C. Plesse). Les grandeurs relatives à la phase solide varient dans l'épaisseur de la lame en raison des variations de concentration en PEDOT; on peut assimiler la lame à un tricouche, les deux couches extérieures riches en PEDOT étant symétriques au repos et d'épaisseur voisine de $30~\mu m$ et la couche centrale pauvre en PEDOT d'épaisseur $190~\mu m$. Au repos, on peut considérer que l'EMITFSI se trouve principalement dans la partie centrale; l'épaisseur des électrodes reste donc très voisine de $30\mu m$ après gonflement. 

D'après N. Festin \cite{Festin}, la masse d'EMITFSI absorbée peut être considérée comme une fonction affine décroissante de la masse de PEDOT lorsque la fraction massique de PEDOT est faible; ce qui permet d'estimer les fractions massiques d'EMITFSI dans chacune des couches. La masse volumique, la charge massique et la fraction volumique au repos de la phase 2, de même que le module d'Young, peuvent dès lors être considérées comme des fonctions créneaux de la coordonnée $z$. Leurs valeurs sont déduites des données fournies par C. Plesse et des valeurs indiquées dans \cite{Festin}, \cite{Festin2013} et \cite{festin2014} et sont récapitulées dans le tableau ci-dessous; les modules d'Young sont estimés à partir des mesures d'A. Fannir \cite{FannirTh} en utilisant une loi de mélange :
\begin{equation}
\begin{tabular}{|c|c|c|c|c|c|c|}
\hline
&$PEDOT$&$EMITFSI$&$\rho_{2}^{0}$&$Z_{2}$&$\phi_{2r}$&$E$\\ \hline
&$\%~massique$&$\%~massique$&$g~cm^{-3}$&$kg~m^{-3}$&$ $ &$MPa$\\ \hline
Bords &$31.9$&$24.3$&$1.02$&$-6.43~10^4$&$0.953$&$137$\\ \hline
Centre &$2.71$&$68.1$&$1.23$&$-2.08~10^{5}$&$0.838$&$15.4$ \\ \hline
\end{tabular}
\end{equation}
On en déduit aussi le module d'Young moyen du matériau $E_{moy}=49,2~MPa$, valeur proche de celle mesurée par N. Festin \cite{Festin2013}. Dans toute la suite, l'indice $b$ fait référence aux électrodes.

\section{Résolution et application au PEDOT}
\medskip
Compte tenu des relations de Maxwell, la loi de Darcy est intégrable pour chaque couche :
\begin{equation}
\left( \frac{1} {\rho_{2}^{0}} - \frac{1}{\rho _{1}^{0}} \right) p - (Z_{1}-Z_{2}) \varphi +\frac{1}{\phi_{2}\rho_{2}^{0}}\frac{3(1+\nu)}{E} p^2=Cte
\end{equation}
Comparons le premier et le dernier terme de cette équation. La valeur maximale de la pression peut être évaluée en utilisant le modèle de la poutre en flexion et la force de blocage $F^{p}=32mN$ mesurée par N. Festin \cite{festin2014} pour $L=3mm$ :
\begin{equation}
|p|_{max} \simeq \frac{\sigma_{xx,max}}{3} \simeq \frac{L F^{p}}{6 \ell e^2}
\end{equation}
Le dernier terme vaut donc au maximum $3,6\%$ du premier et peut être négligé. 

Au repos, la charge électrique massique est nulle en tout point ce qui permet d'écrire :
\begin{equation}
\phi _{1r} (\rho _{1}^{0}Z_{1}-\rho_{2}^{0} Z_{2}) +\rho_{2}^{0} Z_{2} =0
\end{equation}
On en déduit :
\begin{equation}
\phi_{1r} - \phi_{1}= \frac{\varepsilon \varphi"}{\rho _{1}^{0}Z_{1}-\rho_{2}^{0} Z_{2}}= \frac{3 (1 - 2 \nu)}{E} p
\end{equation}

Le système d'équation peut être réécrit sous la forme :
\begin{equation}
\begin{tabular}{l}
$E_{z}=-\frac{d \varphi}{dz}$ \\
$\rho Z =- \varepsilon \frac{d^{2} \varphi}{dz^{2}}$ \\
$D_{z}=- \varepsilon \frac{d \varphi}{dz}$ \\
$\frac{3 (1 - 2 \nu)}{E} p = \frac{\varepsilon }{\rho _{1}^{0}Z_{1}-\rho_{2}^{0} Z_{2}}\frac{d^{2} \varphi}{dz^{2}}$ \\
$\left( \frac{1} {\rho_{2}^{0}} - \frac{1}{\rho _{1}^{0}} \right) p - (Z_{1}-Z_{2}) \varphi \simeq Cte$ \\
\end{tabular}
\end{equation}

Posons :
\begin{equation}
\begin{tabular}{llll}
$\overline{z}=\frac{z}{e} \qquad $&$\overline{\varphi}=\frac{\varphi}{\varphi_{0}} \qquad $&$\overline{\varphi}'=\frac{e}{\varphi_{0}}\frac{d \varphi}{dz} \qquad $&$\overline{\varphi}''=\frac{e^{2}}{\varphi_{0}}\frac{d^{2} \varphi}{dz^{2}}$\\
$\overline{p}=\frac{3 ( 1-2 \nu)}{E_{b}} p\qquad $&$\overline{E}=\frac{e}{\varphi_{0}} E_{z} \qquad $&$\overline{\rho Z}=\frac{\rho Z}{\rho _{1}^{0}Z_{1}} \qquad $&$\overline{D}=\frac{D_{z}}{\rho _{1}^{0}Z_{1}e}$\\
\end{tabular}
\end{equation}
où les ' désignent des dérivées par rapport à $\overline{z}$. On obtient le système d'équations adimensionnées suivant :
\begin{equation}
\begin{tabular}{l}
$\overline{E}=-\overline{\varphi}'$ \\
$\overline{\rho Z} =- A_{0} \overline{\varphi}''$ \\
$\overline{D}=- A_{0} \overline{\varphi}'$ \\
$\overline{p} = A_{0}F_{1}\overline{\varphi}''$ \\
$\overline{\varphi}''-\delta^{2} \overline{\varphi} = Cte$ \\
\end{tabular}
\label{eq-adim}
\end{equation}
où :
\begin{equation}
A_{0} = \frac{\varepsilon \varphi_{0}}{\rho _{1}^{0}Z_{1}e^{2}} \qquad \qquad
F_{1}=\frac{E \phi_{2r}}{E_{b}} \qquad \qquad
\delta = \sqrt{3 (1-2 \nu) \frac{\rho _{1}^{0}Z_{1}e^{2}}{\varepsilon}\frac{\rho_{2}^{0}(Z_{1}-Z_{2})}{E\phi_{2r} \left(1-\frac{\rho_{2}^{0}}{\rho_{1}^{0}}\right)}}
\end{equation}
avec les conditions aux limites :
\begin{equation}
\overline{\varphi}(-1) = 1 \qquad \qquad \overline{\varphi}(1) = -1 \qquad \qquad \overline{\varphi}'(-1) =\overline{\varphi}'(1)
\end{equation}
La solution de l'équation différentielle s'écrit, compte tenu des conditions aux limites :
\begin{equation}
\overline{\varphi} =-\frac{sh(\delta \overline{z})}{sh \delta} \qquad \qquad \overline{\varphi}' =-\frac{\delta }{sh \delta}ch(\delta \overline{z})
 \qquad \qquad \overline{\varphi}'' =-\frac{\delta^{2}}{sh \delta}sh(\delta \overline{z})
\end{equation}

Les équations (\ref{eq-adim}) permettent d'en déduire le profil des différentes grandeurs adimensionnées. D'après (\ref{moment}), le moment fléchissant s'écrit :
\begin{equation}
M^{p} = -\frac{2 \ell E_{b} e^{2}}{1-2 \nu} A_{0} F_{1} \frac{\delta^{2}}{sh \delta} \int_{-1}^{1} sh(\delta \overline{z}) \overline{z} d\overline{z}
=-\frac{4 \ell E_{b} e^{2}}{1-2 \nu} A_{0} F_{1} \left( \frac{\delta}{th \delta}-1 \right)
\end{equation}
Les valeurs de la flèche et de la force de blocage sont alors fournies par les équations (\ref{eq-meca}).

\section{Résultats des simulations numériques}
\medskip

La permittivité diélectrique du tricouche $\varepsilon$ n'a pas été mesurée. Par analogie avec le Nafion \cite{Tixier4}, on peut cependant supposer qu'elle est de l'ordre de $10^{-7}~F~m^{-1}$, d'où l'on déduit $\delta \simeq52300$.
Pour cette valeur de la permittivité et pour $L=3mm$, on obtient une flèche $w=0.72 mm$, une force de blocage $F^{p}=57 mN$ et une déformation $\epsilon_{xx}=2\%$. Ces valeurs sont très proches de celles mesurées par Festin et al \cite{festin2014} : $F^{p}=32 mN$ et $\epsilon_{xx}=2.8\%$.

\begin{figure}[h!]
\begin{center}
\includegraphics[width=0.8\textwidth]{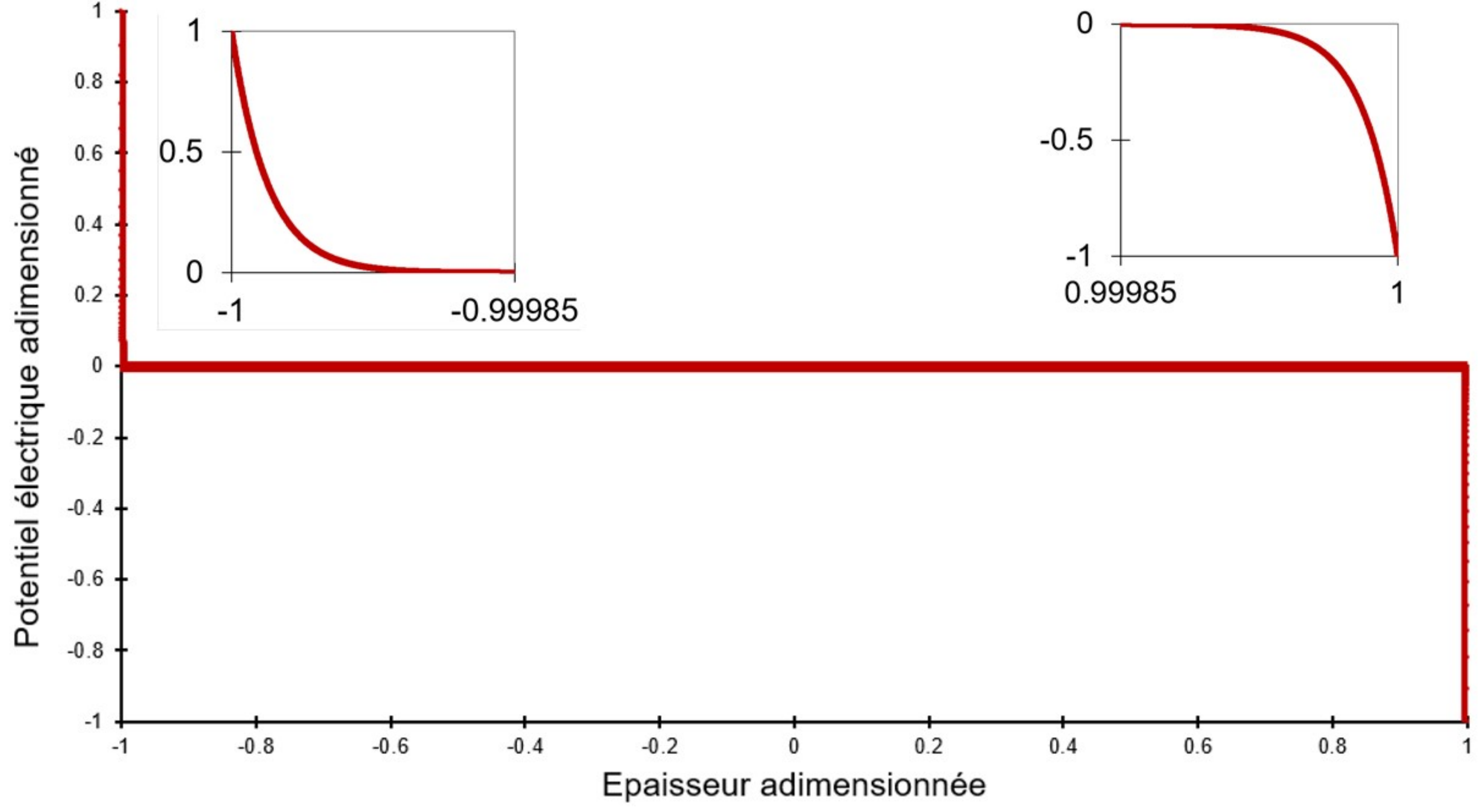}
\caption{Variation du potentiel électrique dans l'épaisseur de la lame}
\label{phi}
\end{center}
\end{figure}

\begin{figure}[h!]
\begin{center}
\includegraphics[width=0.8\textwidth]{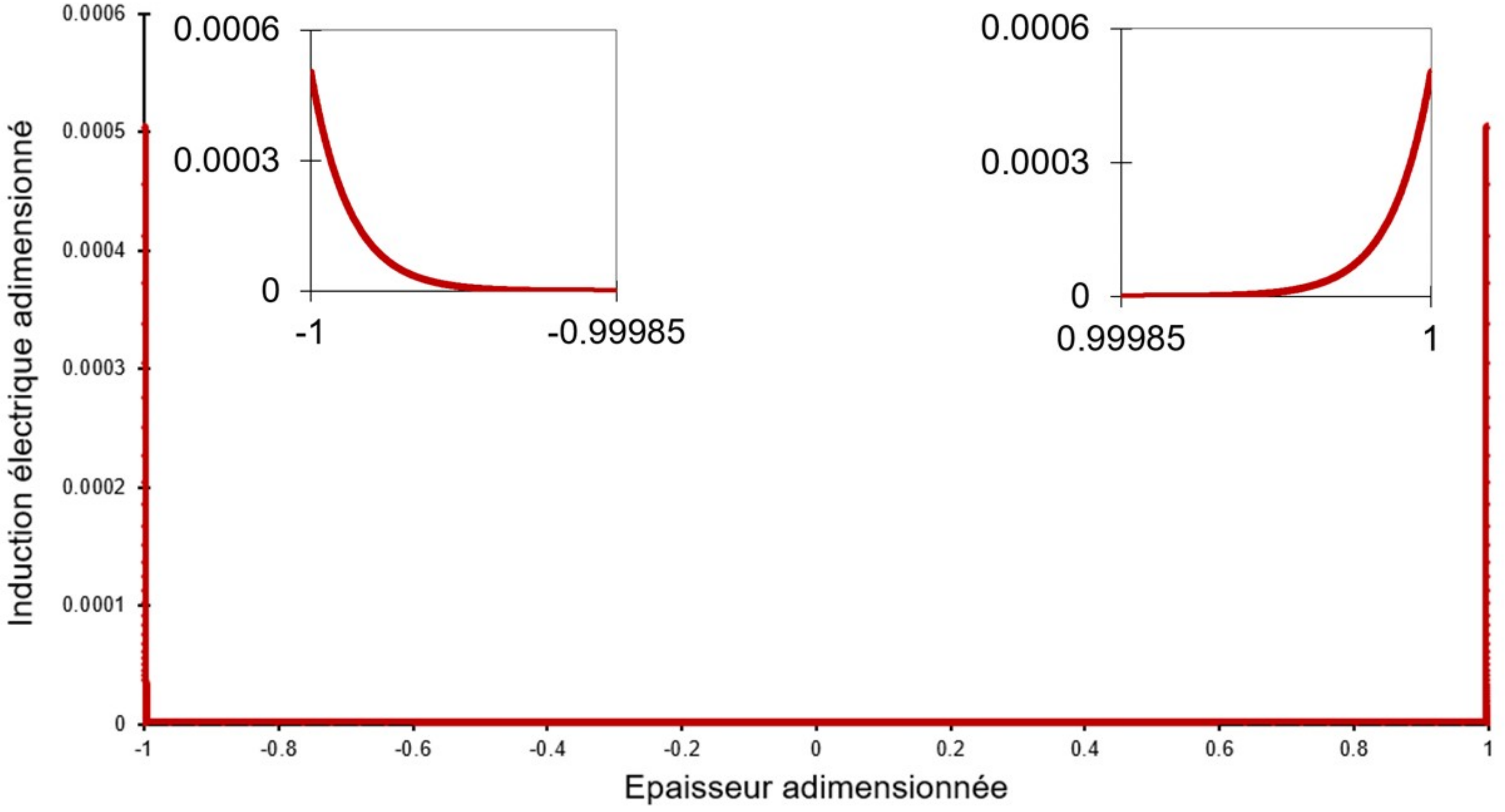}
\caption{Variation de l'induction électrique dans l'épaisseur de la lame}
\label{D}
\end{center}
\end{figure}

\begin{figure}[h!]
\begin{center}
\includegraphics[width=0.8\textwidth]{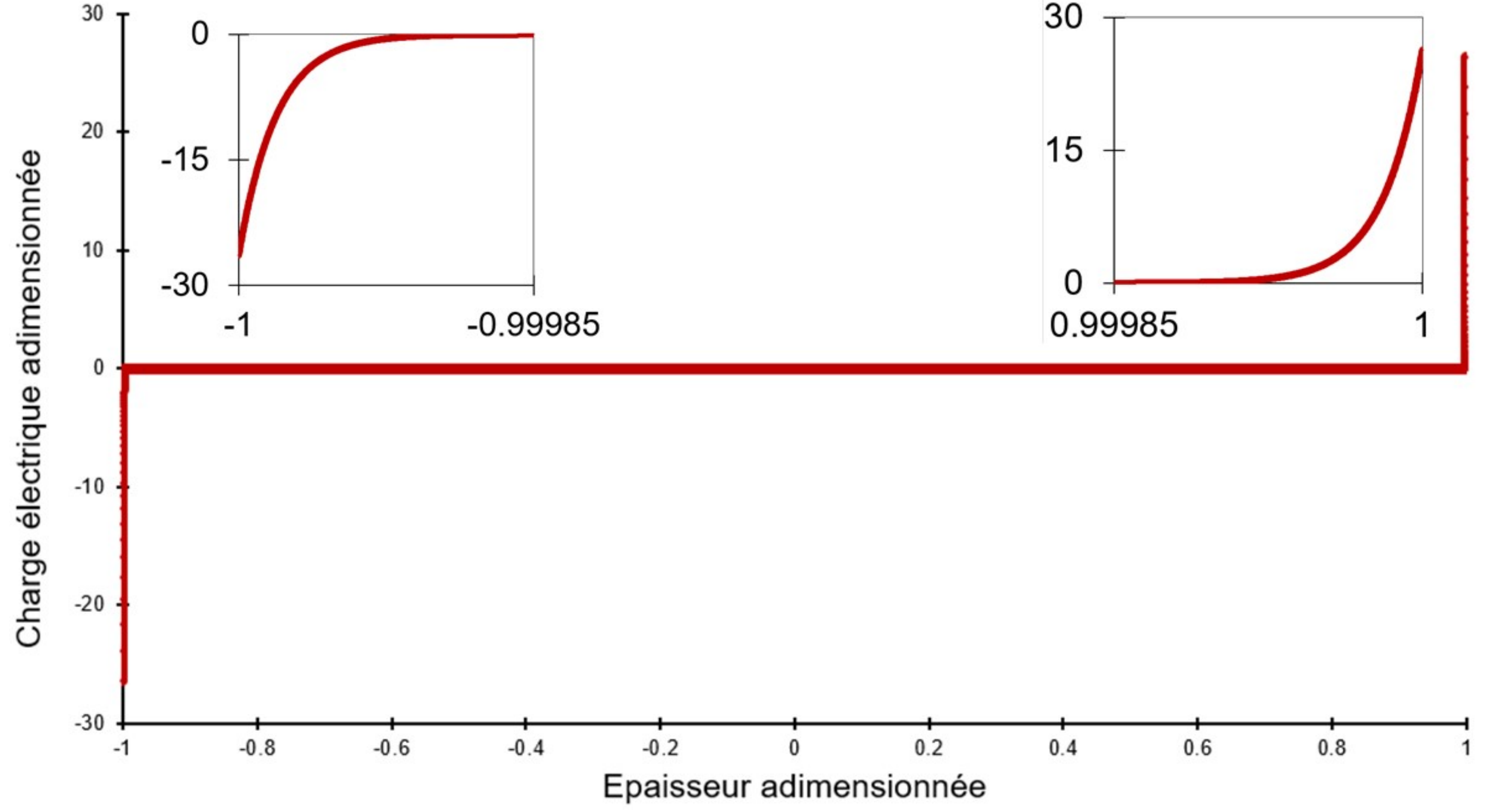}
\caption{Variation de la charge électrique dans l'épaisseur de la lame}
\label{rhoZ}
\end{center}
\end{figure}

\begin{figure}[h!]
\begin{center}
\includegraphics[width=0.8\textwidth]{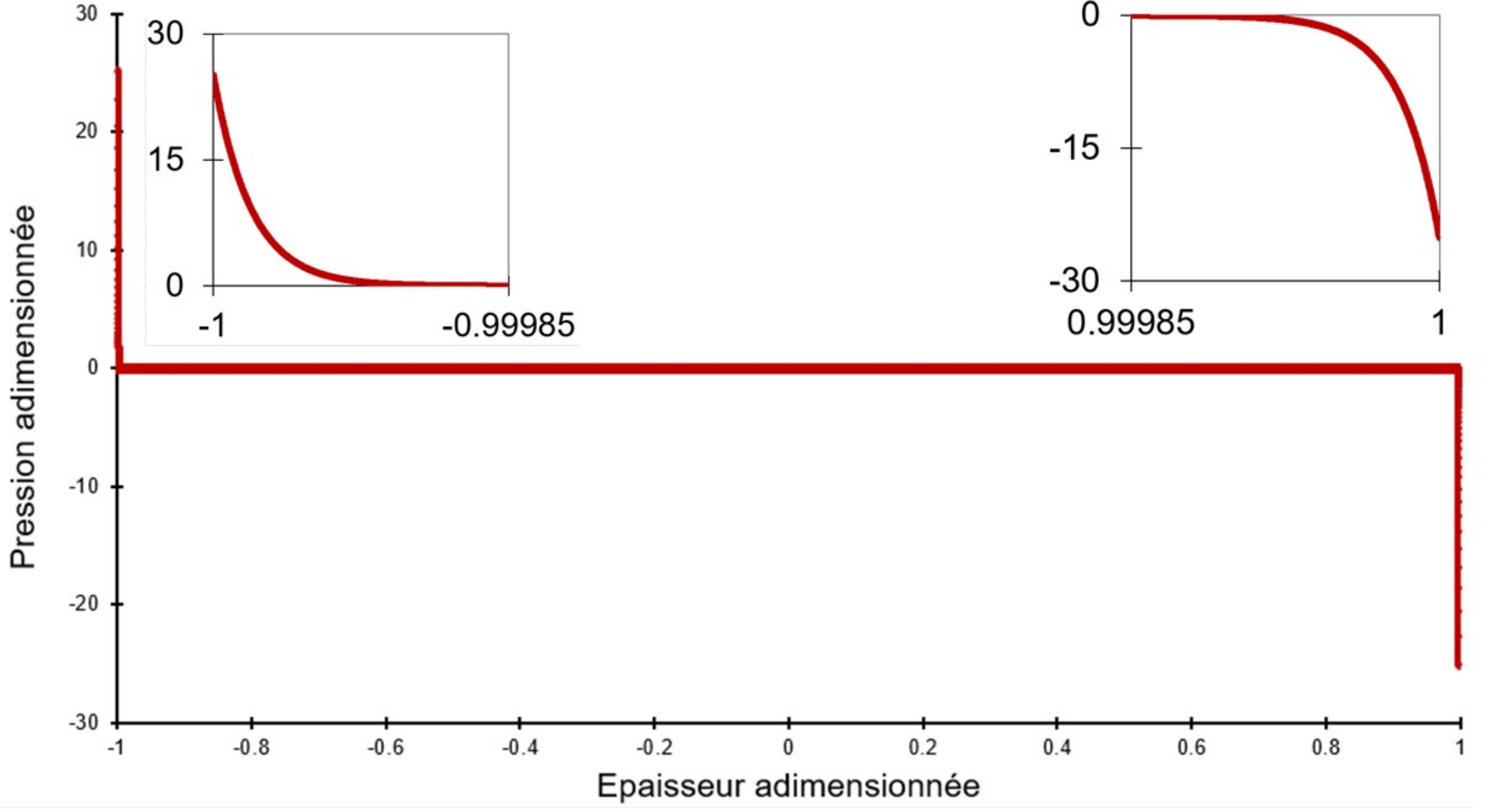}
\caption{Variation de la pression dans l'épaisseur de la lame}
\label{p}
\end{center}
\end{figure}

Les profils des différentes grandeurs sont représentés sur les figures \ref{phi}, \ref{D}, \ref{rhoZ} et \ref{p}. Compte tenu de la valeur de $\delta$, les courbes obtenues sont extrêmement raides au voisinage des bords.
L'induction électrique (figure \ref{D}), de même que la charge électrique (figure \ref{rhoZ}), est nulle sur la quasi totalité de l'épaisseur, ce qui signifie que le matériau se comporte comme un conducteur; le profil du potentiel électrique (figure \ref{phi}) corrobore ce résultat. On vérifie que les cations s'accumulent à la surface de l'électrode supérieure, ce qui provoque une forte augmentation de la contrainte $\sigma_{xx}=-3p$ responsable du fléchissement de la lame (figure \ref{p}); à l'inverse, la surface de l'électrode inférieure devient très électronégative, ce qui signifie que les cations présents au départ ont migré vers la partie centrale.

\section{Conclusion}
\medskip
Nous avons étudié un système complexe constitué de trois polymères interpénétrés saturés d'un liquide ionique. L'un des polymères, le PEDOT, est un polymère électro-actif semi-conducteur. 
Nous avons tout d'abord adapté un modèle développé pour un polymère électro-actif ionique, le Nafion, à ce système. La thermodynamique linéaire des processus irréversibles nous a permis d'obtenir ses lois de comportement.

Nous avons ensuite appliqué ce modèle à la flexion d'une lame encastrée à l'une de ses extrémité et soumise à une différence de potentiel entre ses deux faces dans le cas statique; l'autre extrémité est soit libre, soit soumise à une force de blocage. La composition de la lame n'est pas uniforme : le PEDOT est concentré dans les deux couches externes, alors que la partie centrale, qui fait office de réservoir d'ions, contient la plus grande partie du liquide ionique; on peut donc assimiler ce système à un peudo-tricouche.

Les équations que nous avons obtenues nous ont permis de tracer les profils des différentes grandeurs caractérisant la lame : son potentiel, son induction et sa charge électrique ainsi que sa pression. Les courbes obtenues, constantes dans la partie centrale et très raides au voisinage des bords, montrent que le matériau se comporte comme un conducteur. Nous avons également estimé les valeurs de la déformation et de la force de blocage, qui sont en bon accord avec les données expérimentales.

Nous envisageons par la suite d'étudier l'influence de la géométrie de la lame (longueur, largeur et épaisseur), de même que la variation de la flèche et de la force de blocage en fonction de la différence de potentiel imposée. Nous souhaitons également étudier l'effet inverse, d'abord dans le cas du Nafion, puis dans celui du PEDOT. 

\section{Notations}
\medskip
Les indices $k=1,2,i$ désignent respectivement les cations, la phase solide et les interfaces. Les quantités non indicées sont relatives au matériau complet. $^{s}$ indique la partie symétrique de trace nulle d'un tenseur d'ordre 2 et $\overset{\bullet}{ }$ une dérivée temporelle. L'indice $b$ fait référence aux électrodes et à la partie centrale du tricouche et l'indice $r$ à la lame non déformée.
\newline
\noindent$\overrightarrow{D}$ : induction électrique ;\newline
\noindent$e$ : demi épaisseur de la lame ;\newline
\noindent$E$ : module d'Young ;\newline
\noindent$\overrightarrow{E}$ : champ électrique ;\newline
\noindent$E_{tot}$ ($E_{c}$, $E_{p}$) : énergie totale (cinétique, potentielle) volumique ;\newline
\noindent$\overrightarrow{F^{p}}$ : force de blocage ;\newline
\noindent$\overrightarrow{i}=\overrightarrow{I}-\rho Z \overrightarrow{V}$: courant de diffusion ;\newline
\noindent$\overrightarrow{I}=\sum\limits_{k=1,2,i}\overrightarrow{I_{k}}$ : densité volumique de courant;\newline
\noindent$I^{p}$ : moment quadratique de la poutre par rapport à l'axe $Oy$ ;\newline
\noindent$K$ : perméabilité intrinsèque de la phase solide;\newline
\noindent$l$ : demi largeur de la lame ;\newline
\noindent$L$ : longueur de la lame ;\newline
\noindent$\overrightarrow{M^{p}}$ : moment fléchissant d'axe $Oy$ ; \newline
\noindent$p$ : pression ;\newline
\noindent$\overrightarrow{Q'}=\sum\limits_{k=1,2} \overrightarrow{Q_{k}}, \overrightarrow{Q} = \overrightarrow{Q'}+ \sum\limits_{1,2} \left[ U_{k}(\overrightarrow{V_{k}} - \overrightarrow{V}) -\utilde{\sigma _{k}}.(\overrightarrow{V_{k}}-\overrightarrow{V}) \right]$ : flux de chaleur conductifs ;\newline
\noindent$s$ : production volumique d'entropie ;\newline
\noindent$S$ ($S_{k}$) : entropie volumique totale (du constituant $k$) ;\newline
\noindent$T$ : température absolue ;\newline
\noindent$U$ ($U_{k}$) : énergie interne volumique (du constituant $k$) ;\newline
\noindent$\overrightarrow{V}$ ($\overrightarrow{V_{k}}$): vitesse barycentrique (du constituant $k$) ;\newline
\noindent$w$ : flèche de la poutre ;\newline
\noindent$Z$ ($Z_{k}$) : charge électrique massique du matériau (du constituant $k$) ;\newline
\noindent$\varepsilon $ : permittivité diélectrique ;\newline
\noindent$\utilde{\epsilon }$ : tenseur des déformations ;\newline
\noindent$\eta$ : viscosité dynamique du liquide ;\newline
\noindent$\theta$ : angle de rotation des sections droites de la poutre ;\newline
\noindent$\lambda _{v}$\textit{, }$\mu _{v}$\textit{\ }: coefficients viscoelastiques ;\newline
\noindent$\mu _{k}$ : potentiel chimique massique du constituant $k$ ; \newline
\noindent$\nu$ : coefficient de Poisson du matériau complet ;\newline
\noindent$\rho $ ($\rho _{k},\rho _{k}^{0}$) : masse volumique du matériau (du constituant $k$ ramenée au volume total, du constituant $k$ ramenée au volume de $k$) ;\newline
\noindent$\utilde{\sigma }$ ($\utilde{\sigma_{k}}$) : tenseur des contraintes du matériau (du constituant $k$) ;\newline
\noindent$\utilde{\sigma^{e}}$ ($\utilde{\sigma^{v}}$) : tenseur des contraintes d'équilibre (dynamiques) ;\newline
\noindent$\overrightarrow{\Sigma }$ : flux d'entropie ;\newline
\noindent$\varphi$ ($\varphi_{0}$): potentiel électrique (imposé) ;\newline
\noindent$\phi_{k}$ ($\phi_{2r}$): fraction volumique de la phase $k$ (de la phase 2 au repos) ;\newline

%

\end{document}